\documentclass[sprocl,12pt]{article}
\usepackage{amsmath}

\parskip=\smallskipamount	
\newcommand{\bea}{\begin{eqnarray}}
\newcommand{\eea}{\end{eqnarray}}
\newcommand{\bth}{{\boldsymbol\theta}}

\newcommand{\bpi}{{\boldsymbol\pi}}
\newcommand{\av}[1]{\left\langle #1 \right\rangle}
\newcommand{\rd}[1]{\mathop{\mathrm{d}#1}}

\newcommand{\fract}[2]{{\textstyle\frac{#1}{#2}}}
\newcommand{\grad}{\vec\nabla}
\newcommand{\dtr}{\rd{\vec r}}
\newcommand{\dtrs}{\rd{^2  r}}
\newcommand{\dtrc}{\rd{^3  r}}

\newcommand{\vb}{{\vec b}}
\newcommand{\vk}{{\vec k}}
\newcommand{\vr}{{\vec r}}
\newcommand{\vv}{{\vec v}}

\newcommand{\vcP}{\vec{\mathcal{P}}}

\newcommand{\vA}{{\vec A}}
\newcommand{\ham}{{\hat A_\mu}}
\newcommand{\vB}{{\vec B}}
\newcommand{\vH}{{\vec H}}
\newcommand{\vD}{{\vec D}}
\newcommand{\vE}{{\vec E}}

\newcommand{\pa}{\partial}
\newcommand{\pai}{\partial_i}
\newcommand{\paj}{\partial_j}
\newcommand{\pam}{\partial_\mu}
\newcommand{\pan}{\partial_\nu}

\newcommand{\numeq}[2]{\begin{equation}
#2
\label{#1}
\end{equation}}
\newcommand{\refeq}[1]{(\ref{#1})}

\let\vec\boldsymbol
\let\eps\varepsilon
\let\epsilon\varepsilon
\let\hat\widehat

\begin{document} 

\thispagestyle{empty}

\begin{flushright}
MIT-CTP-3194
\end{flushright}

\vspace{.5cm}

\begin{center}
{\LARGE \bfseries Physical Instances of\\[1ex]
Noncommuting Coordinates}
\vspace{.8cm}  

R. Jackiw
\vspace{0.5cm}

\emph{Center for Theoretical Physics\\ 
Massachusetts Institute of Technology \\
Cambridge MA 02139-4307, USA }

\end{center}

\vspace{.5cm}

\begin{abstract}\noindent
Noncommuting spatial coordinates and fields can be realized in actual physical
situations. Plane wave solutions to noncommuting photodynamics exhibit
violaton of Lorentz invariance (special relativity).
\end{abstract}

\vspace{1.0cm}
\noindent
\rule{6.5cm}{0.4pt}\\
Feza Gursey Institute, Istanbul, Turkey -- June 2001   \\
\llap{``}Symmetry Methods in Physics", Yerevan, Armenia -- July 2001 \\
\llap{``}CPT and Lorentz Symmetry II", Bloomington, IN -- August  2001\\
\llap{``}Particles and Strings", Trento, Italy -- September 2001\\
\llap{``}VIII Adriatic Meeting", Dubrovnik, Croatia -- September 2001

\newpage 

\section{Introduction}

These days, investigators are  probing the validity of Lorentz
invariance (special relativity). This activity is documented by the papers
presented at the Indiana meeting and submitted to the (recently postponed)
Harvard meeting. Experimental and theoretical studies are pursued:
experimentalists measure limits on Lorentz-violating processes; theorists build
plausible Lorentz-violating extensions of the standard model. 

When selecting
Lorentz-violating terms, for possible inclusion in a modified standard model, we 
prefer to use structures that have a preexisting role in physics  or
mathematics. Thus our old proposal to add to the Maxwell Lagrangian the
Lorentz-noninvariant quantity $\frac m2  \int \dtrc \vA\cdot\vB = \frac m2
\int \dtrc
\vA\cdot (\grad\times\vA)$, which leads to birefringence of the vacuum and to
a Faraday-like rotation for the polarization of light propagating through the
vacuum, makes use of the $\int \dtrc \vA\cdot\vB$ quantity, which was
previously known in magnetohydrodynamics as the ``magnetic helicity'', in
fluid mechanics (with the fluid velocity $\vec v$ replacing the electromagnetic
vector potential $\vA$) as the ``kinetic vorticity", and in mathematics as the
``Chern-Simons term". While  the inclusion of this term in an electrodynamical
theory leads to Lorentz, parity, and CTP violation, experiment conclusively
rules out such a modification in Nature~\cite{1}.

Another mechanism for Lorentz-invariance breaking has become the focus of
recent research: the suggestion is made that spatial spatial coordinates need not
commute. While present attention to this idea derives from string theory, we
shall place this mechanism in the more familiar context of quantum
mechanics and quantum field-theory.

Like many interesting quantal ideas, the notion that spatial coordinates may not
commute can be traced to Heisenberg who, in a letter to Peierls, suggested
that a coordinate uncertainty principle may ameliorate the problem of infinite
self-energies. We shall describe later the physical application that Peierls made
with Heisenberg's idea. Evidently, Peierls also described it to Pauli, who told it to
Oppenheimer, who told it to Snyder, who wrote the first paper on the
subject~\cite{2}.

Let us begin with a physical application of the idea that goes back to Peierls. 

\newpage

\section{Noncommutativity in the presence of\protect\\ strong
magnetic fields}
\label{MHD}

\subsection{Particle noncommutativity in the lowest Landau level}

We are interested in a point-particle moving on a plane,
with an external magnetic field $\bf b$ perpendicular to the plane.
The equation for the 2-vector $\vec{r} = (x,y)$ is 
\bea
m \dot{v}^i = \frac{e}{c}\epsilon^{ij}v^jb+f^i(\vec{r})
\label{eq31}
\eea
where $\bf v$ is the velocity $\dot{\vec{r}}$, and $\bf f$ represents
other forces, which we take to be derived from a potential $V$: 
$\vec{f} = - {\grad}V$. Absent additional forces, the quantized theory gives
rise to the well-known Landau levels, with separations $O(b/m)$. The limit
of large
$b$  effectively projects onto the lowest Landau level, and is equivalent to small
$m$. Setting the mass to zero in (\ref{eq31}) leaves a first order equation 
\bea
\dot{r}^i = \frac{c}{eb}\epsilon^{ij}f^j({\bf r})\ .
\label{eq32}
\eea
This may be obtained by taking Poisson brackets of $\bf r$ with 
the Hamiltonian
\bea
H_0 = V
\label{eq33}
\eea
provided the fundamental brackets describe noncommuting coordinates,
\bea
\{ r^i,r^j \} = \frac{c}{eb}\epsilon^{ij}
\label{eq34}
\eea
so that 
\bea
\dot{r}^i = \{ H_0, r^i \} = \{ r^j, r^i \}\partial_j V =
\frac{c}{eb}\epsilon^{ij}f^j({\bf r})\ .
\label{eq35}
\eea

The noncommutative algebra (\ref{eq34}) and the associated dynamics
can be derived in the following manner.
The Lagrangian for the equation of motion (\ref{eq31}) is
\bea
L = \fract12 m {v}^2 + \frac{e}{c}{\vec v} \cdot {\vec A} - V
\label{eq36}
\eea
where we choose the gauge ${\vec A} = (0,bx)$. Setting
$m$ to zero leaves
\bea
L_0 = \frac{eb}{c} x \dot{y} - V(x,y).
\label{eq37}
\eea
which is of the form $p\dot{q} - h(p,q)$, and one sees that
$(\frac{eb}{c}x,y)$ form a canonical pair.  This implies (\ref{eq34}),
and identifies $V$ as the Hamiltonian.  

Finally, we give a canonical derivation of noncommutativity in
the $m \rightarrow 0$ limit, starting with the Hamiltonian
\bea
H = \frac{ {\bf\pi}^2 }{2m} + V\ .
\label{eq38}
\eea
$H$ gives (\ref{eq31}) upon bracketing with $\vec r$ and $\vec\pi$, 
provided the following brackets hold:
\bea
&&\{ r^i, r^j \} = 0 \label{eq39a} \\
&&\{ r^i, \pi^j \} = \delta^{ij} \label{eq39b} \\ 
&&\{ \pi^i,\pi^j \} = -\frac{eb}{c} \epsilon^{ij}\ . \label{eq39c} 
\eea
Here $\bpi$ is the kinematical (noncanonical) momentum,
$m \dot{{\vec r}}$, related to the canonical momentum $\vec p$
by ${\bpi} = {\vec p} - \frac{e}{c}{\vec A}$.

We wish to set $m$ to zero in (\ref{eq38}).
This can only be done provided $\bpi$ vanishes, and
we impose ${\bpi} = 0$ as a constraint.  But according to
(\ref{eq39c}),  the bracket of the constraints 
 is nonzero, and the constraints are recognized to be ``second-class'' in Dirac's
terminology. To proceed with the canonical formalism, we must intro\-duce
Dirac brackets. We omit the details of that technology, but
merely record  the resulting Dirac bracket:
\bea
\{ r^i, r^j \}_D =  \frac{c}{eb}\epsilon^{ij}\ .
\label{eq311}
\eea
In this approach, noncommuting coordinates arise as 
Dirac brackets in a system constrained to lie in the lowest 
Landau level.
Notice that the coordinate noncommutativity is already established at the
classical level in that the Poisson bracket of coordinates is nonvanishing. Later
we shall discuss the quantum version~\cite{3}. 

Peierls observed that when an impurity in the electron system is described
by $V$, one can obtain the first-order energy shift of the lowest Landau level
by taking the coordinates of $(x,y)$ on which $V$ depends to be
noncommuting~\cite{4}. 

A further interesting subject, which is not discussed here, concerns the behavior
of the wave function in the phase-space reductive, $m\to0$, limit that projects
onto the lowest Landau level. Before the reduction, the wave function is a
normalized expression depending on the two coordinates. After the reduction,
the wave function can depend only on one coordinate, because the other is a
conjugate variable. How all this comes about is explained in the
literature~\cite{3}.

\subsection{Field noncommutativity in the lowest Landau level}

The above demonstrates that spatial coordinates of particles in an intense
magnetic field do not (Poisson) commute. But we are interested in fields. To
find an example of noncommuting fields, we turn to the equations of a
charged fluid, moving on a plane in an external magnetic field perpendicular
to plane. The fluid is described by a 
 density 
$\rho$ and velocity $\vv$, both defined on the two-dimensional plane. A
mass parameter
$m$ is introduced for dimensional reasons, so that the mass density is
$m\rho$. The fields $\rho$ and $\vv$ are 
functions of $t$ and $\vec r$ and give an Eulerian description
of the fluid.  The equations that are satisfied are the
continuity equation 
\bea
\dot{\rho} + {\grad} \cdot (\rho{\vec v}) = 0
\label{eq312}
\eea
which expresses matter conservation, 
and the Euler equation 
\bea
m \dot{v}^i + m{\vec v} \cdot {\grad} v^i = 
\frac{e}{c}\epsilon^{ij}v^jb + f^i 
\label{eq313}
\eea
which is the force equation. 
Here $f^i$ describes additional forces, e.g., 
$-\frac{1}{\rho} {\grad} P$ where $P$ is pressure.
We shall take the force to be derived from a potential of the 
form 
\bea
{\vec f}({\vec r}) = -{\grad} 
\frac{\delta}{\delta\rho({\vec r})} \int \dtr V.
\label{eq314}
\eea
(For isentropic systems, the pressure is only a function of
$\rho$; (\ref{eq314}) holds with V a function of $\rho$,
related to the pressure by $P(\rho) = \rho V^{\prime}(\rho)- V(\rho)$.
Here we allow more general dependence of $V$ on $\rho$
(e.g. nonlocality or dependence on derivatives of $\rho$)
and also translation noninvariant, explicit dependence on $\vec r$~\cite{5}.)

Equations (\ref{eq312})--(\ref{eq314}) follow by bracketing 
$\rho$ and $\vec \pi$ with the Hamiltonian
\bea
H= \int\dtrs  \Bigl( \rho \frac{ {\bf\pi}^2 }{2m} + V \Bigr)
\label{eq315}
\eea
provided that fundamental brackets are taken as
\begin{align}
\{ \rho({\vec r}), \rho({\vec r}^{\prime}) \} &= 0 \label{eq316a} \\
\{ \pi({\vec r}), \rho({\vec r}^{\prime}) \} &= 
{\grad} \delta({\vec r} - {\vec r}^{\prime}) \label{eq316b} \\
\{ \pi^i({\vec r}), \pi^j({\vec r}^{\prime}) \} &=
-\epsilon^{ij}\frac{1}{\rho(\vr)}\Bigl(m \omega({\vec r}) + 
\frac{eb}{c}\Bigr) \delta({\vec r}-{\vec r}^{\prime}) \label{eq316c}
\end{align}
where $\epsilon^{ij}\omega({\vec r})$ is the vorticity 
$\partial_iv^j -\partial_jv^i$, and ${\bpi} = m {\vec v}$.

We now consider a strong magnetic field and take the limit 
$m\rightarrow 0$, which is equivalent to large $b$. 
Equations (\ref{eq313}) and (\ref{eq314})
reduce to
\bea
v^i = -\frac{c}{eb}\epsilon^{ij} \frac{\partial}{\partial r^j}
\frac{\delta}{\delta\rho({\vec r})} \int\dtrs  V\ .
\label{eq317}
\eea
Combining this with the continuity equation (\ref{eq312})
gives the equation for the density ``in the lowest Landau level'':
\begin{equation}
\dot{\rho}({\vec r}) = \frac {c}{eb}\frac{\partial}{\partial r^i}
\rho({\vec r}) \epsilon^{ij}\frac{\partial}{\partial r^j}
\frac{\delta}{\delta \rho({\vec r})}\int\dtrs  V
\label{eq318}
\end{equation}
(For the right-hand side not to vanish, $V$ must not be solely a 
function of $\rho$.)

The equation of motion (\ref{eq318}) can be obtained by 
bracketing with the Hamiltonian
\bea
H_0 = \int\dtrs  V
\label{eq319}
\eea
provided the charge density bracket is nonvanishing, showing 
noncommutativity of the $\rho$'s~\cite{6}: 
\bea
\{ \rho({\vec r}), \rho({\vec r}^{\prime}) \} = 
-\frac{c}{eb}\epsilon^{ij}\partial_i\rho({\vec r})
\partial_j\delta( {\vec r}-{\vec r}^{\prime} )\ .
\label{eq320}
\eea

$H_0$ and this bracket may be obtained from (\ref{eq315}) and
(\ref{eq316a}) -- (\ref{eq316c}) with the same Dirac procedure presented for
the particle case: We wish to set $m$ to zero in (\ref{eq315});
this is possible only if $\bpi$ is constrained to vanish.
But the bracket of the $\bpi$'s \refeq{eq316c} is nonvanishing, even at $m=0$,
because $b \ne 0$. Thus at $m=0$ we are dealing with a second-class
constraint which leads to a nonvanishing Dirac  bracket of densities  as in
(\ref{eq320}):
\bea
 \{ \rho({\vec r}),\rho({\vec r}^{\prime}) \}_D =  
 - \frac{c}{eb} \epsilon^{ij}\partial_i \rho({\vec r})
\partial_j\delta({\vec r} - {\vec r}^{\prime}) \ .
\label{eq323}
\eea

The $\rho$-bracket \refeq{eq320}, \refeq{eq323} enjoys a more appealing
expression in momentum space.  Upon defining 
\bea
{\tilde{\rho}}({\vec p}) = \int\dtrs  e^{i{\vec p}\cdot{\vec r}}
\rho({\vec r})
\label{eq324}
\eea
we find
\bea
\{ {\tilde{\rho}}({\vec p}), {\tilde{\rho}}({\vec q}) \} =
-\frac{c}{eb}\epsilon^{ij}p^iq^j
\tilde{\rho}({\vec p} + {\vec q}).
\label{eq325}
\eea
  
The form of the charge density bracket (\ref{eq320}), (\ref{eq323}), 
(\ref{eq325}) can be understood by reference to the particle 
substructure for the fluid. Take
\bea
\rho({\vec r}) = \sum_n \delta({\vec r} - {\vec r}_n) 
\label{eq326}
\eea
where $n$ labels the individual particles.  When the coordinates of 
each particle satisfy the nonvanishing bracket (\ref{eq34}), (\ref{eq311}), the
$\{ \rho({\vec r}), \rho({\vec r}^{\prime}) \}$ bracket takes the form
(\ref{eq320}), (\ref{eq323}), (\ref{eq325}).

\subsection{Quantization}
\label{quantize}

Quantization before the reduction to the lowest Landau level is
straightforward. For the particle case (\ref{eq39a})--(\ref{eq39c}) 
and for the 
fluid case (\ref{eq316a})--(\ref{eq316c})
we replace brackets with $i/\hbar$
times commutators.  After reduction to the lowest Landau level 
we do the same for the particle case thereby arriving at the 
``Peierls substitution,'' which (as mentioned previously) states that the effect of 
an impurity [$V$ in (\ref{eq36})] on the lowest Landau energy
level can be evaluated to lowest order by viewing the $(x,y)$ arguments 
of $V$ as noncommuting variables~\cite{4}.

For the fluid, quantization presents a choice.
On the one hand, we can simply promote the 
bracket (\ref{eq320}), (\ref{eq323}), (\ref{eq325}) to a 
commutator by multiplying by $i/\hbar$.
\begin{align}
 [ \rho({\vec r}),\rho({\vec r}^{\prime}) ] &= 
i\hbar\frac{c}{eb} \epsilon^{ij}\partial_i \rho({\vec r}^{\prime})
\partial_j\delta({\vec r} - {\vec r}^{\prime}) \label{eq327a} \\
 \left[ \tilde{\rho}({\vec p}), \tilde\rho({\vec q}) \right] &=
i \hbar \frac{c}{eb} \epsilon^{ij} p^i q^j \tilde\rho({\vec p} + {\vec q})
\label{eq327b}
\end{align}

Alternatively we can adopt the expression (\ref{eq326}), for the operator 
$\rho({\vec r})$, where the ${\vec r}_n$ now satisfy the
noncommutative algebra
\bea
\left[  r_n^i,  r_{n^{\prime}}^j \right] = -i\hbar \frac{c}{eb} 
\epsilon^{ij}\delta_{nn^{\prime}}
\label{eq328}
\eea
and calculate the $ \rho$ commutator as a derived
quantity.

However, once ${\vec r}_n$ is a noncommuting operator, functions
of $ {\vec r}_n$, even $\delta-$functions,
have to be ordered. We choose the Weyl ordering,  which is equivalent to 
defining the Fourier transform as 
\bea
\tilde{\rho}({\vec p}) = \sum_n e^{i{\vec p} \cdot {\vec r}_n}\ .
\label{eq329}
\eea
With the help of (\ref{eq328}) and the Baker-Hausdorff lemma, we
arrive at the ``trigonometric algebra''
\bea
[\tilde{\rho}({\vec p}), \tilde{\rho}({\vec q})]=
2i \sin \Bigl( \frac{\hbar c}{2eb} \epsilon^{ij}p^iq^j \Bigr)
\tilde{\rho}({\vec p}+ {\vec q})\ .
\label{eq330}
\eea
This reduces to (\ref{eq327b}) for small $\hbar$.

This form for the  commutator, (\ref{eq330}), 
is connected to a Moyal star product in
the following fashion.  For an arbitrary $c$-number function $f({\vec r})$
define
\bea
\langle f\rangle  = \int\dtrs  \rho({\vec r})f({\vec r}) = 
\frac{1}{(2\pi)^2}\int \rd{^2 p} \tilde{\rho}({\vec p}) \tilde{f}(-{\vec p})\ .
\label{eq331}
\eea
Multiplying (\ref{eq330}) by 
$\tilde{f}(-{\vec p})\tilde{g}(-{\vec q})$ and integrating
gives
\begin{gather}
[\langle f\rangle , \langle g\rangle ] = \langle h\rangle 
\label{eq332}\\
\intertext{with}
h(\vec{r}) = (f\star g)(\vec{r}) - (g\star f)(\vec{r})
\label{eq333}
\end{gather}
where the ``$\star $'' product is defined as 
\bea
(f\star g)(\vec{r}) = e^{\frac{i}{2}\frac{\hbar c}{eb} 
\epsilon^{ij} \partial_i \partial_j^{\prime}}
f(\vec{r})g(\vec{r}^{\prime})|_{ \vec{r}^{\prime} = \vec{r}}.
\label{eq334}
\eea
Note however that only the commutator is mapped into the star 
commutator.  The product $\langle f\rangle \langle g\rangle $ is not equal to
$\langle f\star g\rangle $.

The lack of consilience between (\ref{eq327b}) and (\ref{eq330})
is an instance of the Groenwald-VanHove theorem which establishes 
the impossibility of taking over into quantum mechanics all classical
brackets~\cite{7}. Equations (\ref{eq328})--(\ref{eq334}) explicitly
exhibit the physical occurrence of the star product for fields in
a strong magnetic background. 

\section{Various algebras}

Before proceeding with  our construction  of a noncommutative Maxwell field
theory, let us summarize here the various (nontrivial) algebras that we have
encountered in the above development.

The fluid velocity algebra \refeq{eq316c} at $b=0$ and $m=1$ reads in any
spatial dimension
\numeq{ne37}{
\{v^i(\vr), v^j(\vr')\} = -\frac1{\rho(\vr)} \bigl(
\pai v^j(\vr) -\paj v^i(\vr)\bigr) \delta(\vr-\vr')\ .
}
This was first given by Landau~\cite{8}. In spite of the awkward appearance,
the algebra in fact takes a familiar form when we define the momentum density
$\vcP = \rho\vv$, and use \refeq{eq316a}, \refeq{eq316b} for the
$\rho$~brackets. Then \refeq{ne37}, with \refeq{eq316a} and
\refeq{eq316b} implies 
\numeq{ne38}{
\{\mathcal P^i (\vr), \mathcal P^j (\vr')\} = \Bigl(\mathcal P^j (\vr)
\frac{\pa}{\pa r^i} +  \mathcal P^i (\vr') \frac\pa{\pa r^j}\Bigr) 
\delta(\vr-\vr')\ . 
}
This is the usual momentum density algebra, which also describes
diffeomorphisms of space in the following fashion. If an infinitesimal coordinate
transformation is given by 
\numeq{ne39}{
\delta r^i = -f^i(\vr)
}
we define the average $\av f$ of $f^i$ by integrating with $\mathcal P^i$
\numeq{ne40}{
\av f \equiv \int \dtr f^i (\vr) P^i(\vr)
}
then \refeq{ne38} has the consequence that for two such functions~$f$ and~$g$
we have
\numeq{ne41}{
\{\av f, \av g\} = -\av h
}
where $h$ is the Lie bracket of $f$ and $g$:
\numeq{ne42}{
h^i =  g^j \paj f^i - f^j\paj g^i \ .
}

By scaling $\rho$ the noncommutative density algebra \refeq{eq320},
\refeq{eq323} may be presented as 
\numeq{ne43}{
\{\rho(\vr), \rho(\vr')\} = \eps^{ij} \pai \rho(\vr)\paj \delta(\vr-\vr')\ . 
}
This intrinsically two-dimensional structure is the  area-preserving algebra,
studied by Arnold~\cite{9}. Area-preserving coordinate transformations
(volume preserving in arbitrary dimensionality) possess unit Jacobian. For the
infinitesimal form of the transformation \refeq{ne39} this means that $f^i$ is
transverse: $\pai f^i =0$. Therefore, in two dimensions, an area-preserving
transformation is generated by a scalar:
\numeq{ne44}{
f^i = \eps^{ij} \paj f\ . 
}
When an average $\av f$ is defined by 
\numeq{ne45}{
\av f = \int \dtrs f(\vr) \rho(\vr) 
}
equation \refeq{ne43} again implies \refeq{ne41}, but now we have
\numeq{ne46}{
h = \eps^{ij} \pai f \paj g
}
which also follows from \refeq{ne42} when all three functions take the
form~\refeq{ne44}.

Finally the algebra  \refeq{eq330} 
\[
\{\tilde{\rho}({\vec p}), \tilde{\rho}({\vec q})\}=
-\frac2\hbar \sin \Bigl( \frac{\hbar c}{2eb} \eps^{ij}p^i q^j \Bigr)
\tilde{\rho}({\vec p}+ {\vec q}) 
\]
which also leads to the Moyal-star product \refeq{eq334} for averages
\refeq{ne45}, is called a trigonometric algebra, which was introduced by
D.~Fairlie, P.~Fletcher, and C.~Zachos~\cite{10}.

\section{Noncommutative electrodynamics}

Stimulated by the occurrence of the star product in the discussion of charged
fluids in an intense magnetic field, we abstract the idea and use it in the new
setting of noncommutative Maxwell theory. This theory is described by the
vector potential~$\ham$ (the caret denotes noncommuting quantities)
and the  theory is built on a gauge-invariance principle. Gauge transformations
act on $\ham$ according to
\numeq{ne47}{
\ham \to \ham^\lambda = (e^{i\lambda})^{-1} \star (\ham -i\pam) \star 
(e^{i\lambda}) \ .
}
The star ($\star$) product of two quantities is defined by 
\numeq{ne48}{
(O_1 \star O_2)(\vr) = e^{\frac i2 \theta^{\mu\nu} \frac\pa{\pa r^\mu} \frac
\pa{\pa r^{\prime\nu}}} O_1(\vr) O_2(\vr') \bigr|_{\vr=\vr'}
}
and we take $\theta^{\mu\nu}$ to have no time components ($\theta^{0i} = 0$,
$\theta^{ij}=\eps^{ijk}\theta^k$). The  field strength $\hat F_{\mu\nu}$ is
constructed from
$\ham$ in a manner such that the gauge transformation \refeq{ne47} effects a
covariant transformation:
\numeq{ne49}{
\hat F_{\mu\nu}\to \hat F_{\mu\nu}^\lambda = 
(e^{i\lambda})^{-1} \star F_{\mu\nu} \star 
(e^{i\lambda}) \ .
}
This requirement is met, provided $\hat F_{\mu\nu}$ is given by
\numeq{ne50}{
\hat F_{\mu\nu} = \pam \hat A_\nu - \pan \hat A_\mu - i [\hat A_\mu,\hat A_\nu]_\star
}
where $[\hat A_\mu,\hat A_\nu]_\star = \hat A_\mu \star \hat A_\nu - \hat A_\nu \star \hat A_\mu$.
Finally, the action is taken to be 
\numeq{ne51}{
\begin{aligned}
\hat I &= -\fract14 \int \rd{^4 x} \hat F^{\mu\nu} \star \hat F_{\mu\nu}\\
    &= -\fract14 \int \rd{^4 x} \hat F^{\mu\nu}  \hat F_{\mu\nu}\ .
\end{aligned}
}

One would like to find the equations of motion, calculate physically interesting
quantities, and compare them to corresponding quantities in the Maxwell
theory. In this way one could assess the effect of noncommutativity and
perhaps place experimental limits on it. However, a problem arises: local
quantities in noncommutative electrodynamics are gauge variant and no
invariant meaning can be assigned to their profiles. Nonlocal, integrated,
expressions can be gauge invariant,  (for example, the action \refeq{ne51} is
gauge invariant) but in the ordinary Maxwell theory we deal with local
quantities (like profiles of electromagnetic waves) and we would like to
compare these classical local disturbances to corresponding quantities in the
noncommutative theory. 

A way out of this difficulty is provided by Seiberg and Witten's observation that
the  noncommuting gauge theory may be equivalently described by a
commuting gauge theory that is formulated in terms of ordinary (not star)
products of a commuting vector potential $A_\mu$, together with an explicit
dependence on $\theta^{\alpha\beta}$, which acts as a constant ``background''.
This equivalence is established by expressing the noncommuting vector
potential $\ham$ as a function of $A_\mu$ and $\theta^{\alpha\beta}$ that
solves the Seiberg-Witten equation~\cite{11}
\numeq{ne52}{
\frac{\pa\ham}{\pa \theta^{\alpha\beta}} = 
-\fract18 \bigl\{  
\hat A_\alpha, \pa_\beta \ham + \hat F_{\beta\mu}
\bigr\}_\star -(\alpha \leftrightarrow \beta)
}
where the bracketed expression denotes the ``star'' anticommutator. Solutions
of this equation are expressed in terms of $\theta^{\alpha\beta}$ and the
``initial condition'' $\ham\bigr|_{\theta^{\alpha\beta}=0}$; the latter quantity
being just the commuting~$A_\mu$.

We work to lowest order in $\theta$ and find
\numeq{ne53}{
\ham = A_\mu - \fract12 \theta^{\alpha\beta} 
  A_\alpha  \bigl(\pa_\beta A_\mu + F_{\beta\mu}
\bigr) \ .
}
The noncommuting action, expressed in terms of the commuting quantities
$A_\mu$, $F_{\mu\nu} = \pam A_\nu - \pan A_\mu$, and
$\theta^{\alpha\beta}$,  now reads~\cite{12} 
\numeq{ne54}{
\hat I = -\fract14 \int \rd{^4 x} \Bigl(
(1- \fract12 \theta^{\alpha\beta} F_{\alpha\beta}) F^{\mu\nu} F_{\mu\nu}
+ 2\theta^{\alpha\beta} F_{\mu\alpha} F_{\nu\beta} F^{\mu\nu}
\Bigr)\ .
}
This is gauge invariant in the conventional sense, and from the equations of
motion that are implied by $\hat I$ we can determine the gauge-invariant
electric ($E^i = F^{i0}$) and magnetic fields ($B^i = -\eps^{ijk} F_{jk}$).

These fields satisfy the equations, which maintain a Maxwell form.
\begin{subequations}\label{ne55}
\begin{align} 
\frac1c \frac\pa{\pa t} \vB + \grad \times \vE &= 0 \label {ne55a}\\
\grad\cdot \vec B & = 0 \label{ne55b}
\end{align}
\end{subequations}
\vspace*{-\bigskipamount}
\begin{subequations}\label{ne56}
\begin{align} 
\frac1c \frac\pa{\pa t} \vD - \grad \times \vH &= 0 \label {ne56a}\\
\grad\cdot \vD & = 0 \label{ne56b}
\end{align}
\end{subequations}
The first set \refeq{ne55} reflects the gauge invariance of the system, namely,
that $\vE$ ant $\vB$ are given in terms of  potentials. The second set
\refeq{ne56} is a consequence of the nonlinear dynamics implied by
\refeq{ne54}. The constitutive relations relating $\vD$ and $\vH$ to $\vE$ and
$\vB$ follow from
\refeq{ne54}:
\numeq{ne57}{
\begin{aligned}
\vD &= (1-\vec\theta \cdot \vB)\vE + (\vec\theta\cdot\vE)\vB + 
             (\vE\cdot\vB)\vec\theta\\
\vH  &= (1-\vec\theta \cdot \vB)\vB - (\vec\theta\cdot\vE)\vE + 
             \fract12(\vE^2 - \vB^2)\vec\theta \ .
\end{aligned}
}
Note that parity is preserved -- coordinate reflection leaves the constant vector
$\vec\theta$ unchanged, hence $\vec\theta \cdot \vB$ transforms as a scalar
field and $\vec\theta \cdot \vE$ as a pseudoscalar field. Similarly,
$(\vE\cdot\vB)\vec\theta$ behaves as a vector field, while $(\vE^2 -
\vB^2)\vec\theta$ as a pseudovector.

We seek plane-wave solutions to \refeq{ne55}--\refeq{ne57} --- functions of
$\omega t -\vk\cdot\vr$ --- keeping terms to lowest order in $\theta$. Such
solutions indeed exist  provided the dispersion relation, relating $\vk$
and $\omega$, takes the following form.  In the absence of an external
magnetic field the dispersion relation is conventional, $\omega=ck$. However,
plane wave solutions to our system of equations exist even in the presence of a
constant background magnetic induction~$\vb$. Then the dispersion relation is
modified to
\numeq{ne58}{
\omega = ck(1-\vec\theta_T\cdot\vb_T)
}
where $\vec\theta_T$ and $\vb_T$ are components transverse to $\vk$, the
direction of propagation $\vk\cdot\vec\theta_T = \vk\cdot\vb_T = 0$~\cite{6}.

The result \refeq{ne58} puts into evidence an explicit violation of Lorentz
invariance. Conservation of parity, which we remarked on previously, ensures
that both  polarizations travel at the same velocity, which generically 
differs from $c$ by the factor $(1-\vec\theta_T\cdot\vb_T)$, and there is
no Faraday rotation. 
Let us also observe that the effective Lagrange density in \refeq{ne54}
possesses two interaction terms proportional to $\theta$, 
with definite numerical constants.  Owing to the freedom
of rescaling $\theta$, only their ratio is significant.   It is
straightforward to verify that if the ratio is different 
from what is written in \refeq{ne54}, the two linear polarizations
travel at different velocities. Thus the non-commutative theory is 
unique in affecting the two polarizations equally, at least to 
$O(\theta)$.

The change in velocity for motion relative to an external magnetic 
induction $\vb $ allows searching for the effect with a 
Michelson-Morley experiment. In a conventional apparatus with
two legs of length $\ell_1$ and $\ell_2$ at right angles to each
other, a light beam of wavelength $\lambda$ is split in two, and
one ray travels along $\vb $ (where there is no effect), while
the other, perpendicular to $\vb $, feels the change of velocity
and interferes with the the first. After rotating the apparatus by 
$90^{\circ}$, the interference pattern will shift by 
$2(\ell_1 + \ell_2){\bth}_T \cdot \vb_T/ \lambda$ fringes.
Taking light in the visible range, $\lambda \sim 10^{-5}$cm, 
a field strength $b \sim 1$ tesla, and using the current bound on
$\theta \le (10$TeV$)^{-2}$ obtained in \cite{13}, one finds that
a length $\ell_1 + \ell_2 \ge 10^{18}$~cm~$\sim 1$~parsec
would be required for a shift of one fringe.  Galactic magnetic 
fields are neither that strong nor coherent over such large distances,
so another experimental setting needs to found to test for
non-commutativity.

Finally, what about Heisenberg's  intuition that noncommuting coordinates will
ammeliorate divergences in relativistic field theory? It turns out that that is
indeed true as far as ultraviolet divergences are concerned. However, novel
infrared divergences appear, so the problem of divergences remains, albeit in
another form. Indeed, these infrared effects associated with noncommutative
coordinates provide another obstacle to physical applicatons of this idea.

%
%

\end{document}